\documentclass{aa}

\usepackage[varg]{txfonts}
\usepackage{siunitx}
\usepackage[version=4]{mhchem}
\usepackage{multirow}
\usepackage{graphicx}
\usepackage[caption=false]{subfig}
\usepackage[strict]{changepage}
\usepackage{xcolor}
\usepackage{ulem}
\usepackage[colorlinks=true,citecolor=blue]{hyperref}
\usepackage{float}
\usepackage{amssymb}
\usepackage{amsmath}
\usepackage{lscape}
\usepackage{mathtools}
\usepackage{natbib}
\usepackage{pdflscape}
\usepackage{gensymb}

\defcitealias{zhou2023}{Paper~I}   

\DeclareSIUnit\jansky{Jy}

\newcolumntype{L}[1]{>{\raggedright\let\newline\\\arraybackslash\hspace{0pt}}m{#1}}
\newcolumntype{C}[1]{>{\centering\let\newline\\\arraybackslash\hspace{0pt}}m{#1}}
\newcolumntype{R}[1]{>{\raggedleft\let\newline\\\arraybackslash\hspace{0pt}}m{#1}}

\begin{document}

\title{The dependence of black hole formation in open clusters on the cluster formation process}
\author{Jian-Wen Zhou\inst{\ref{inst1}} 
}
\institute{
Max-Planck-Institut f\"{u}r Radioastronomie, Auf dem H\"{u}gel 69, 53121 Bonn, Germany \label{inst1} \\
\email{jwzhou@mpifr-bonn.mpg.de}
}

\date{Accepted XXX. Received YYY; in original form ZZZ}

\abstract
{
We performed N-body simulations of both individual cluster evolution and subcluster coalescence, demonstrating that cluster evolution and its outcomes strongly depend on the cluster formation process through comparisons of different gas expulsion modes and formation channels. The evolution of star clusters is significantly shaped by the gas expulsion mode, with faster expulsion producing greater mass loss. A broader degeneracy exists among initial cluster mass, gas expulsion timescale, and formation channel (monolithic vs. coalescence), which manifests in both evolutionary pathways and black hole production. In individual cluster simulations, slower gas expulsion enables progressively lower-mass clusters to retain central black holes within the tidal radius. As the gas expulsion mode transitions from fast to moderate to slow, the fraction of high-velocity stars decreases. 
Variations in gas expulsion mode and formation channel ultimately influence the stellar velocity distribution (within the tidal radius), and thus the expansion speed, which governs both cluster mass loss and black hole retention. Slowly expanding clusters are more likely to retain black holes and multiple systems, making them prime candidates for black hole searches with {\it Gaia}. Our results highlight the crucial influence of early gas expulsion and cluster formation mechanisms on the dynamical evolution of star clusters and black hole production. These factors should be carefully incorporated into the initial conditions of N-body simulations, which necessarily rely on input from the star formation community.
}


\titlerunning{}
\authorrunning{}

\maketitle

\section{Introduction}

Recent proper motion data from {\it Gaia}’s Data Release 3 (DR3; \citet{Gaia2023-674}) has led to the discovery of two dormant black holes (BHs, i.e. {\it Gaia} BH1 and {\it Gaia} BH2) in BH–star binary systems within the Galactic field \citep{Chakrabarti2023-166,El-Badry2023-518,El-Badry2023-521,Tanikawa2023-946}.  
As part of the validation process for the upcoming fourth Gaia data release (DR4), \citet{Gaia2024-686} analyzed preliminary astrometric binary solutions and identified a black hole with a mass of $\sim$33 $M_{\odot}$ in a binary system, now referred to as {\it Gaia} BH3. 
Following the detection of {\it Gaia} BH1 and {\it Gaia} BH2, discussions and hypotheses emerged regarding the origins of these systems. The formation histories of {\it Gaia} BH1 and {\it Gaia} BH2 are challenging to reconcile with isolated binary (IB) evolution, as their orbital properties do not align with the expected outcomes of common envelope evolution \citep{El-Badry2023-521,El-Badry2023-518}. Their near-solar metallicities and inferred orbits within the Galactic plane suggest that they were likely formed in open clusters \citep{Rastello2023-526,DiCarlo2024-965,Marin2024-688,Tanikawa2024-527}.

Young and open star clusters host most massive stars, which are the precursors of compact objects. Consequently, the majority of black holes in the Milky Way likely spent their early years in these clusters, engaging in dynamic interactions.
N-body simulations are powerful tools for studying star cluster dynamics and black hole formation within clusters. Physically, the initial conditions of N-body simulations are determined by the star cluster formation process. Given the significant progress in observational studies of massive star and star cluster formation \citep{Motte2018-56,Beuther2025-63}, these observational results should be translated into the initial conditions for star cluster dynamical simulations (see Appendix.\ref{nbody} for more details). 
This work also examines how initial conditions and the cluster formation process influence the evolutionary outcomes of star clusters.

\begin{table*}
\centering
\caption{N-body simulations modeling both the coalescence of multiple subclusters and the evolution of individual clusters under different gas expulsion modes. More details can refer to Sec.\ref{simu} and Appendix.\ref{nbody}.
"Time" is the time when the simulation is terminated. The cluster may or may not have dissolved by this time. 
}
\label{tab1}
\begin{tabular}{ccccc}
\hline
Cases	&	Type & Mass (M$_{\odot}$)	&	Time (Myr)	&	Dissolved	\\
10000-fast	&	individual	&	10000	&	263	&	yes	\\
3000-fast	&	individual	&	3000	&	100	&	yes	\\
3000-moderate1	&	individual	&	3000	&	441	&	no	\\
3000-moderate2	&	individual	&	3000	&	256.25	&	yes	\\
3000-slow	&	individual	&	3000	&	950.25	&	yes	\\
1000-fast	&	individual	&	1000	&	67.5	&	yes	\\
1000-moderate1	&	individual	&	1000	&	145.5	&	yes	\\
1000-moderate2	&	individual	&	1000	&	78	&	yes	\\
1000-slow	&	individual	&	1000	&	533.5	&	yes	\\
300-fast	&	individual	&	300	&	69	&	yes	\\
300-moderate1	&	individual	&	300	&	100	&	yes	\\
300-moderate2	&	individual	&	300	&	45.25	&	yes	\\
300-slow	&	individual	&	300	&	346.5	&	yes	\\
NGC6334-fast	&	coalescence	&	2207	&	614.25	&	yes	\\
NGC6334-fast-vd	&	coalescence	&	2207	&	463.75	&	yes	\\
\hline
\label{all}
\end{tabular}
\end{table*}

\section{N-body simulations}\label{simu}

A portion of the N-body simulations for star clusters presented in \citet{Zhou2024-691-293,Zhou2024sub-merger,Zhou2024sub-mM} are summarized in Table.\ref{all}. The detailed simulation setups have also been thoroughly documented in these studies. 
For the convenience of the reader, these setups are summarized again in Appendix.\ref{nbody}.
These simulations were previously employed to explain the observed physical parameters of open clusters. In this work, however, we focus on the final products of these simulations, i.e., black holes. Overall, the simulations can be divided into two categories: the cluster coalescence simulations \citep{Zhou2024sub-merger} and the individual cluster simulations \citep{Zhou2024-691-293}, i.e. "individual" and "coalescence" in Table.\ref{all}. 
As described in Appendix.\ref{nbody}, for individual clusters, we simulated four gas expulsion modes (fast, moderate2/mod2, moderate1/mod1, and slow), characterized by gas expulsion timescales of $\tau_g$, 2$\tau_g$, 5$\tau_g$, and 10$\tau_g$, respectively. To mitigate the stochasticity of the simulation results, we further refined the mass grid and simulated additional clusters using the same setups. All new simulations (i.e., except those listed in Table.\ref{all}) were evolved until 100 Myr in order to reduce computational costs. Finally, the simulated cluster masses are [300, 562, 1000, 1333, 1778, 2371, 3000, 5623, 10000] M$_{\odot}$, corresponding to [10$^{2.5}$, 10$^{2.75}$, 10$^{3}$, 10$^{3.125}$, 10$^{3.25}$, 10$^{3.375}$, 10$^{3.5}$, 10$^{3.75}$, 10$^{4}$] M$_{\odot}$. For the cluster complex "NGC6334" in the coalescence simulations, each subcluster adopts the fast gas expulsion mode. As described in Appendix.\ref{nbody}, we considered two scenarios: (i) all subclusters are initially at rest and located in the same plane ("-fast"), and (ii) the subclusters are spatially separated and possess velocity differences ("-fast-vd").

\section{Results and discussion}\label{result}

\subsection{Cluster evolution}

\begin{figure}
\centering
\includegraphics[width=0.475\textwidth]{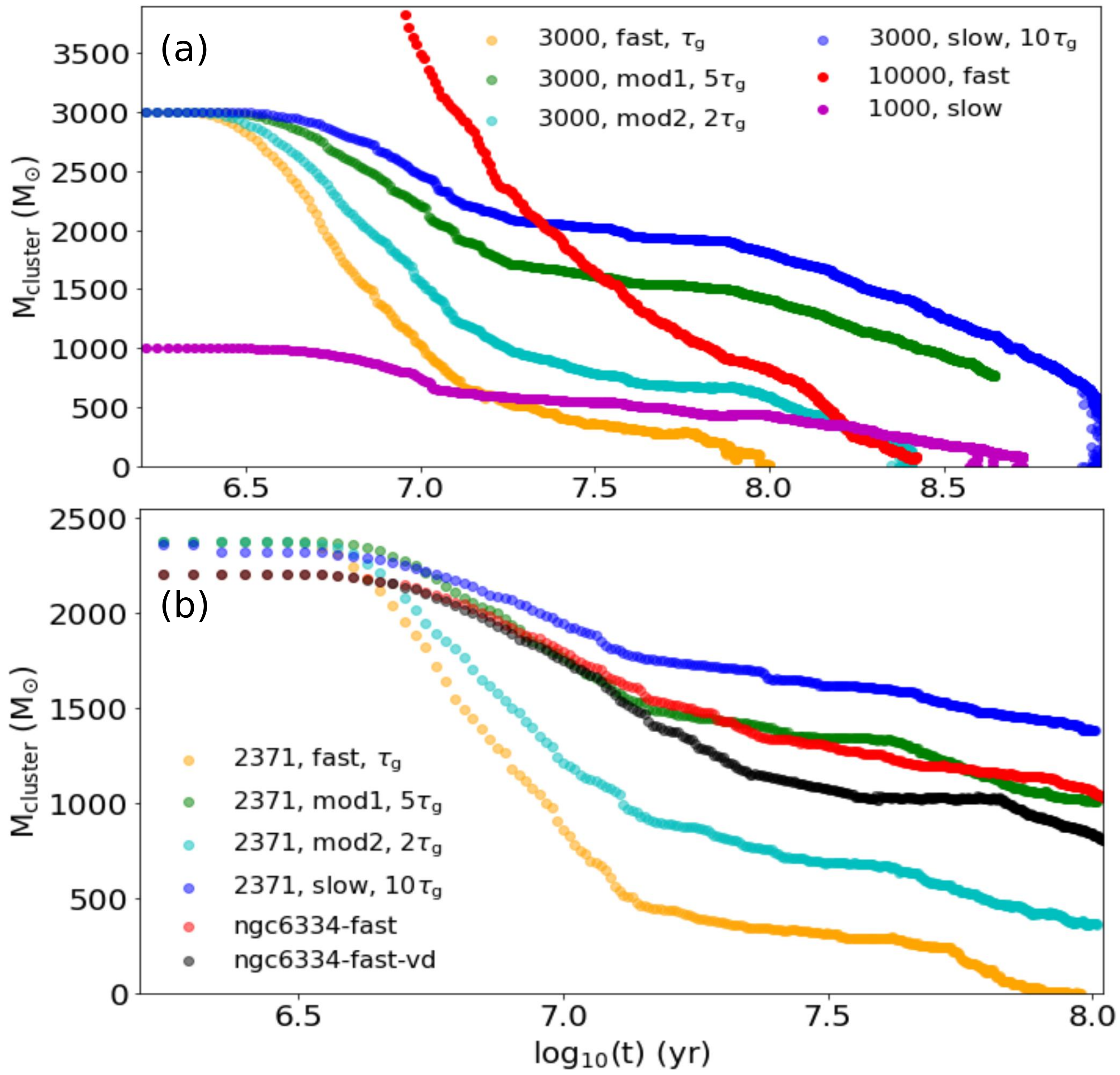}
\caption{The evolution of cluster mass over time under different gas expulsion modes. In panel (b), The total mass of the cluster complex ``NGC6334'' is 2207 M$_{\odot}$, close to the single cluster with mass 2371 M$_{\odot}$. Here we only consider the total cluster mass within the tidal radius.}
\label{Mt}
\end{figure}

The evolution of clusters under different simulation scenarios has been extensively analyzed in \citet{Zhou2024-691-293,Zhou2024sub-merger,Zhou2024sub-mM}. Here we summarize several key conclusions.  
For individual cluster simulations, the extent of cluster mass loss is strongly governed by the gas expulsion mode: the faster the gas expulsion, the greater the mass loss. However, there exists a degeneracy between cluster mass and gas expulsion timescale. 
For example, a massive cluster undergoing fast gas expulsion expands rapidly and loses mass at a high rate, whereas a lower-mass cluster experiencing slow gas expulsion evolves more gradually, resulting in slower mass loss. Consequently, clusters with different initial masses and gas expulsion timescales may display similar evolutionary outcomes. As shown in Fig.\ref{Mt}(a), a 1000~M$_{\odot}$ cluster with slow gas expulsion evolves similarly to a 3000~M$_{\odot}$ cluster with moderate gas expulsion. Conversely, a 10,000~M$_{\odot}$ cluster with fast gas expulsion exhibits evolutionary behavior comparable to that of a 3000~M$_{\odot}$ cluster with moderate gas expulsion. 
In cluster coalescence simulations, the spatial distribution, relative velocities, mass distribution, and gas expulsion modes of subclusters jointly influence both the dynamics of the merging process and the stability of the final product. 
As shown in Fig.\ref{Mt}(b), although each subcluster undergoes fast gas expulsion, the evolution of the merged cluster closely resembles that of a single cluster with moderate gas expulsion in the individual cluster simulations. 
These findings highlight a degeneracy among cluster mass, gas expulsion mode, and the cluster formation channel (i.e., monolithic versus coalescence).

\subsection{Black hole production}\label{bhf}

\begin{figure*}
\centering
\includegraphics[width=0.95\textwidth]{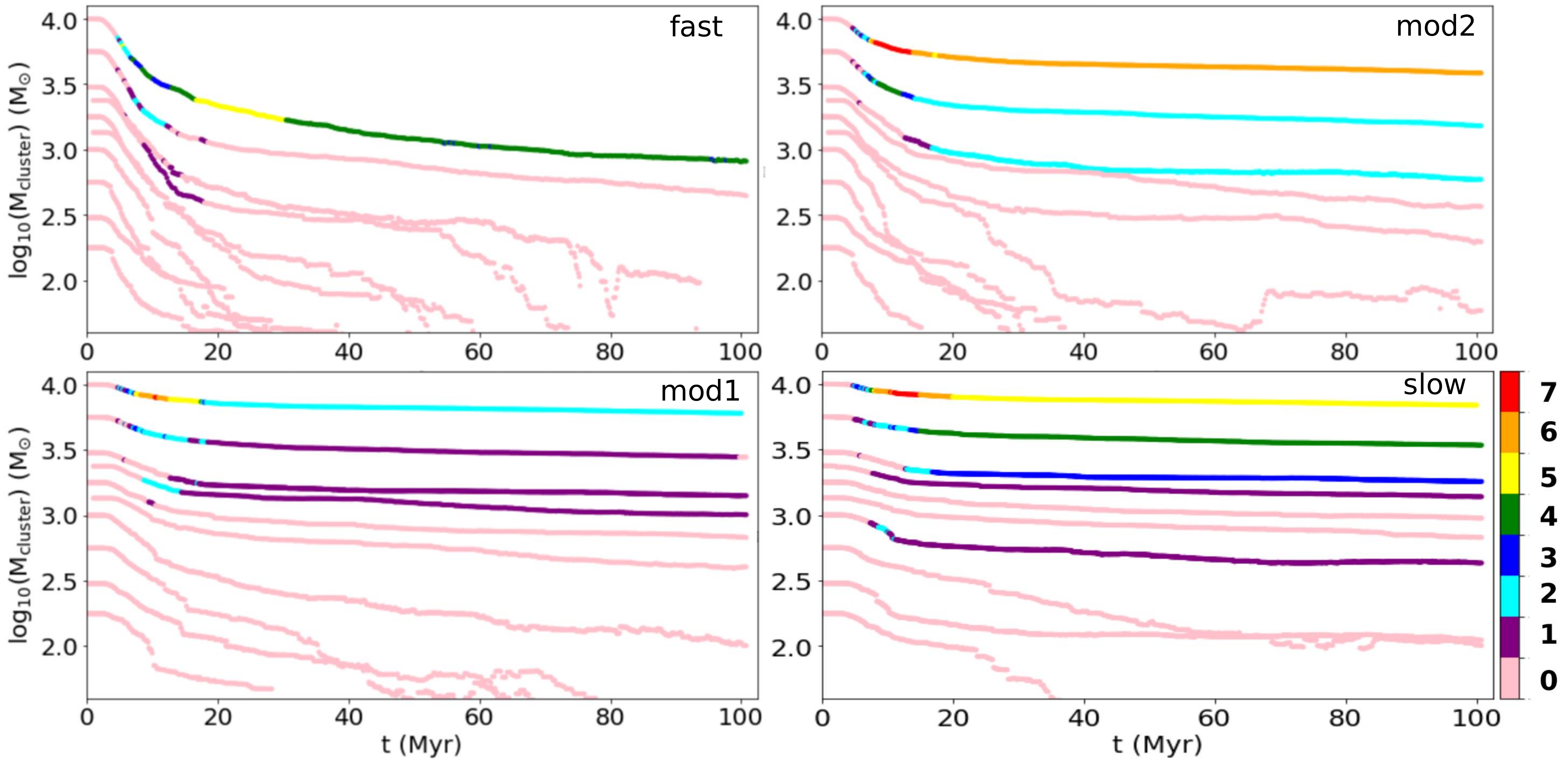}
\caption{The evolution of cluster mass over time under different gas expulsion modes. The color of the line reflects the number of black holes contained in a cluster of a given mass at a certain age. Here, we only consider the total cluster mass and the number of black holes within the tidal radius. For each panel, from bottom to top, the cluster masses are [300, 562, 1000, 1333, 1778, 2371, 3000, 5623, 10000] M$_{\odot}$.}
\label{num}
\end{figure*}

\begin{figure}
\centering
\includegraphics[width=0.475\textwidth]{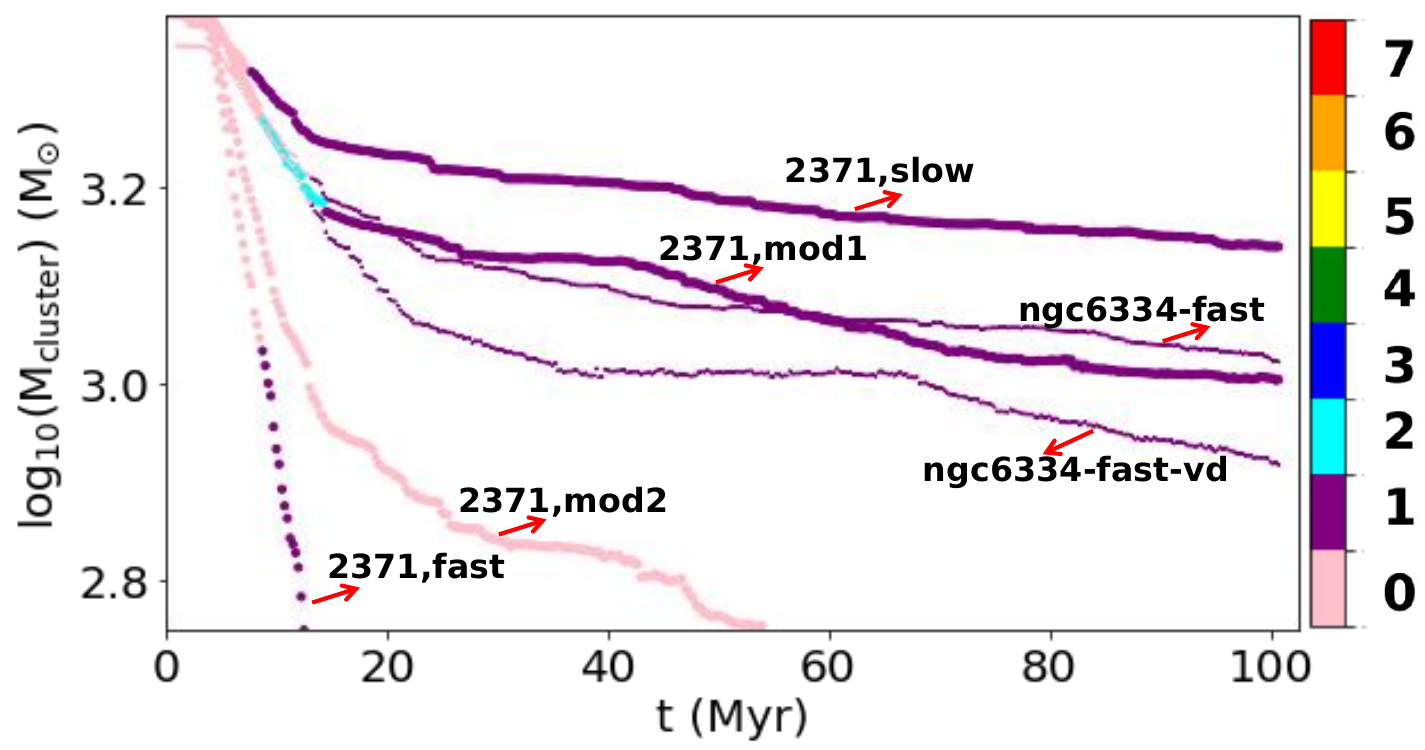}
\caption{Same as Fig.\ref{num}, but for the cluster complex ``NGC6334'' and the 2371 M$_{\odot}$ cluster.}
\label{num1}
\end{figure}

\begin{figure*}[htbp!]
\centering
\includegraphics[width=0.8\textwidth]{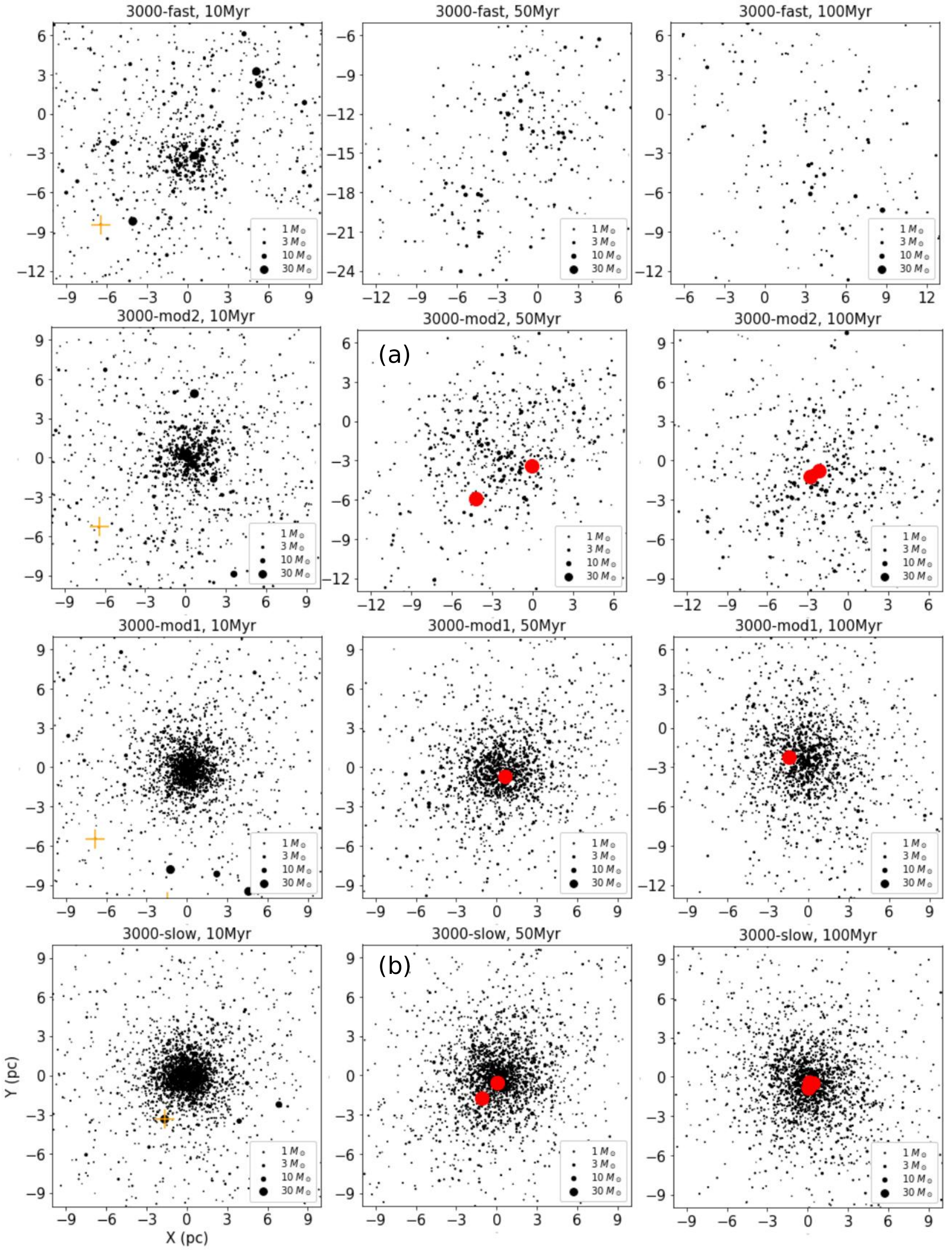}
\caption{The evolution of the 3000 M$_{\odot}$ cluster under different gas expulsion modes. Snapshots projected on the $XY$ plane (20 pc $\times$ 20 pc around the cluster center) at ages 10 Myr, 50 Myr, and 100 Myr are shown here. The size of the black dots reflects the stellar mass (linear scale). Neutron stars and black holes are represented by orange plus signs and red dots, respectively, with the symbol sizes being purely illustrative.
The central regions of the star clusters in Fig.\ref{3000}(a) and (b) contain 2 and 3 black holes, respectively. In panel (b), two of the three black holes are too close to be distinguished on the map.}
\label{3000}
\end{figure*}

\begin{figure}
\centering
\includegraphics[width=0.45\textwidth]{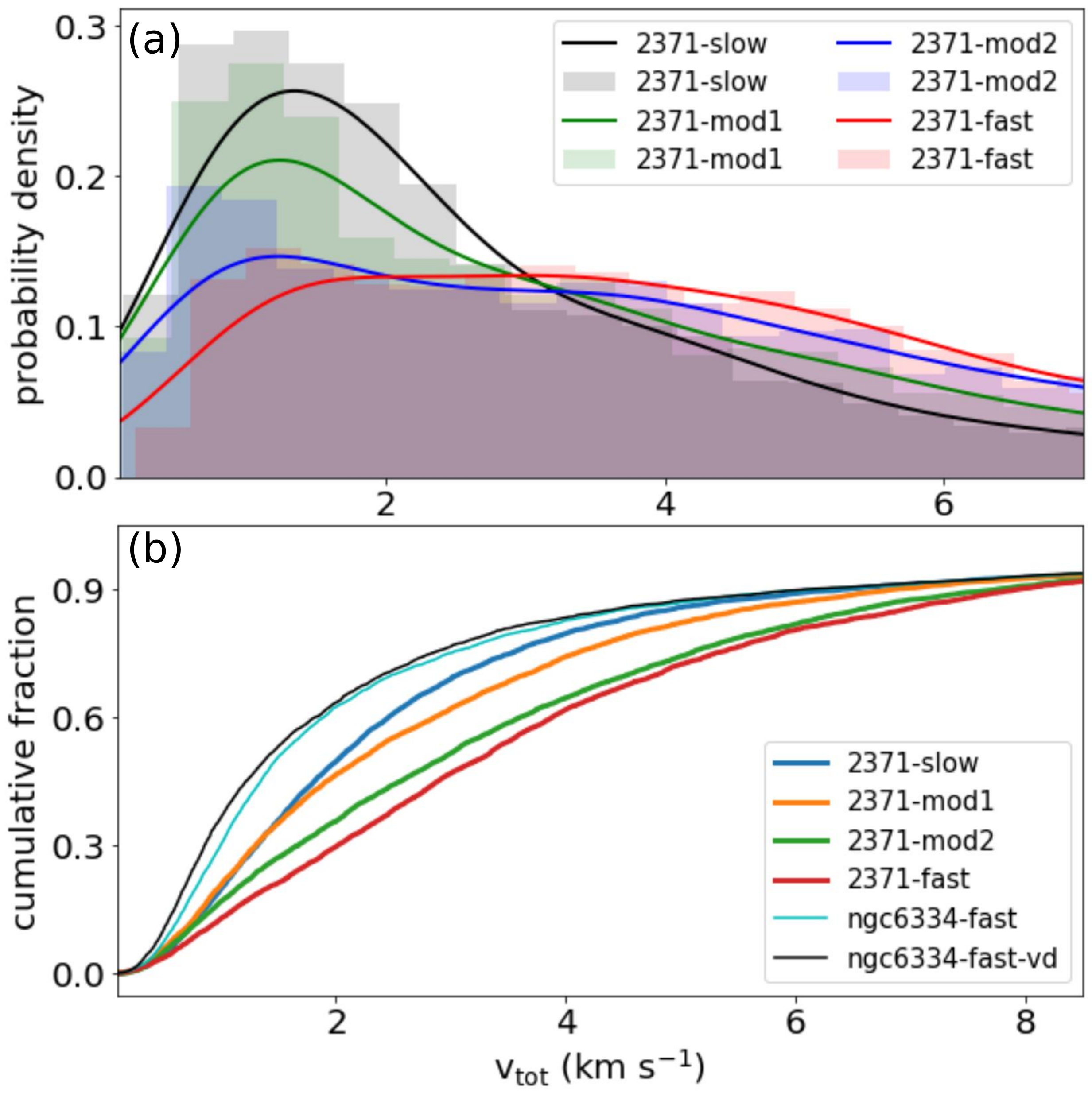}
\caption{Velocity distribution of stars within the cluster complex ``NGC6334'' and the 2371 M$_{\odot}$ cluster under different gas expulsion modes at 20 Myr.
(a) Probability density function (PDF) of stellar velocities, estimated using the kernel density estimation (KDE) method to provide a smooth representation of the distribution; (b) Cumulative fraction shows the proportion of stars with velocities below a given value.
Here we do not distinguish between binaries and single stars; instead, we calculate the total velocity of each star ($v_{\rm tot}$) in the cluster and examine their distribution to roughly characterize the expansion state of the cluster. At 50 Myr, the velocity distributions remain unchanged, indicating that the binary fraction does not significantly affect the statistical results here.}
\label{vt}
\end{figure}

\begin{figure}
\centering
\includegraphics[width=0.45\textwidth]{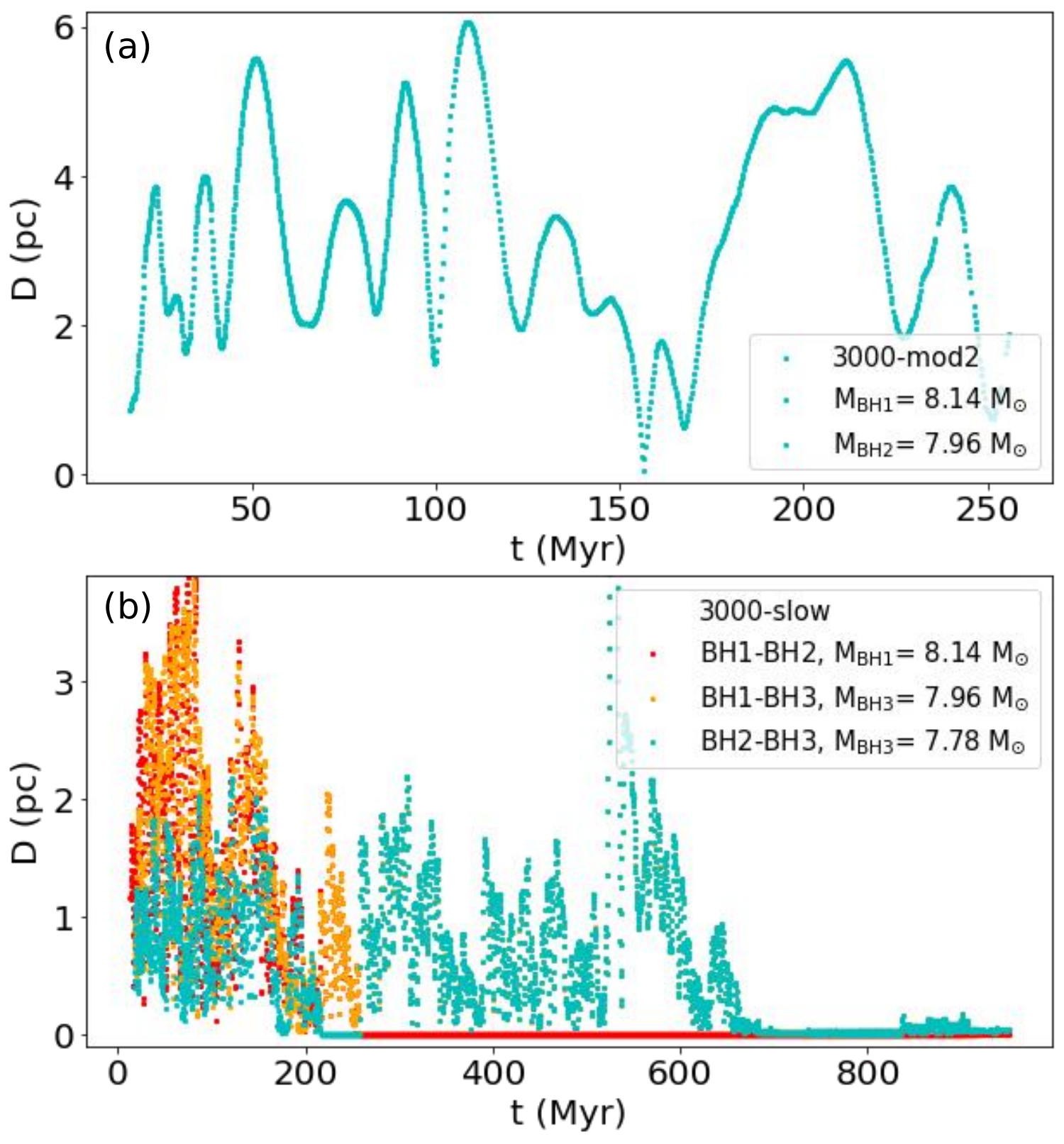}
\caption{The temporal evolution of the distance between each pair of black holes. Panels (a) and (b) correspond to black hole pairs in panels (a) and (b) of Fig.\ref{3000}, respectively.}
\label{dis}
\end{figure}

The degeneracy between cluster mass and the initial gas expulsion mode is also reflected in the eventual black hole production of star clusters, as shown in Fig.\ref{num}. For individual cluster simulations, under fast gas expulsion, only the most massive cluster (10,000~M$_{\odot}$) can retain black holes within the tidal radius for extended periods. As the gas expulsion mode transitions from fast to moderate and then to slow, progressively lower-mass clusters are able to retain black holes within the tidal radius, and the number of retained black holes also increases. When the gas expulsion mode changes from ``fast'' ($\tau_g$) to ``moderate2'' (2$\tau_g$), the timescale itself changes only slightly; however, the subsequent cluster evolution and final outcome differ substantially. This underscores the importance of carefully constraining the initial conditions of N-body simulations of cluster evolution (see Sec.\ref{expulsion} for further discussion).

In cluster coalescence simulations (Fig.\ref{num1}), although each initial subcluster undergoes fast gas expulsion, the merged cluster with a initial mass of 2207~M$_{\odot}$ can still retain a black hole within the tidal radius. This again reflects the degeneracy between cluster mass and the cluster formation mechanism. 
However, coalescence simulations are far more complex than individual cluster simulations, involving a rich parameter space that remains to be explored. 
This work only illustrates the significant differences between cluster coalescence and the evolution of a single cluster by comparing a cluster complex (``NGC6334'') with a single cluster (2371~M$_{\odot}$) of similar mass, rather than providing a detailed discussion of cluster coalescence itself.

Fig.\ref{vt} shows the velocity distribution of stars in the cluster complex ``NGC6334'' and in the 2371~M$_{\odot}$ cluster under different gas expulsion modes. 
As the gas expulsion mode transitions from fast to moderate to slow, the fraction of high-velocity stars in the cluster decreases. In contrast, for the cluster complex, despite each subcluster undergoing fast gas expulsion, the velocity distribution of the coalesced cluster is significantly lower overall. The stellar velocity distribution of a star cluster reflects its expansion speed, which plays a crucial role in determining whether massive stars and their remnants remain bound. Variations in cluster mass, gas expulsion mode, and formation mechanism ultimately influence the cluster expansion speed, which in turn affects both the mass-loss rate and the rate of black hole formation.

In Fig.\ref{3000}, closely spaced black holes are shown. Their ability to form multiple systems (binary or triple) also depends on the dynamical state of the cluster. 
The initial gas expulsion mode is critical to the cluster’s dynamical evolution and has a strong impact on its lifetime. Black holes in slowly expanding clusters experience a more stable dynamical environment and thus have significantly more time to evolve within the system. 
The number of central black holes in Fig.\ref{3000}(a) and (b) is 2 and 3, respectively. Fig.\ref{dis} illustrates the time evolution of the separations between each pair of black holes. The three black holes in Fig.\ref{3000}(b) eventually form a triple system, whereas the two black holes in Fig.\ref{3000}(a) fail to form a binary even by the time the cluster dissolves.


\subsection{Gas expulsion mode}\label{expulsion}

In the above analysis, we demonstrated the crucial influence of the gas expulsion mode and cluster formation mechanism on the dynamical evolution of star clusters and black hole production. Therefore, these factors should be carefully incorporated into the initial conditions of N-body simulations.

Giant molecular clouds (GMCs) are widely recognized as the primary gas reservoirs fueling star formation and serving as the birthplaces of nearly all stars.
Within these clouds, clumps are dense, localized structures that act as the main sites of star formation \citep{Urquhart2018-473,Urquhart2022-510}. 
The process of star cluster formation in molecular clouds involves several stages: the formation of clumps in diffuse gas, the evolution of these clumps, the emergence of embedded clusters within them, and the subsequent evolution of the embedded clusters alongside clump dispersal driven by feedback \citep{Krumholz2019-57,Krause2020-216,Zhou2022-514,Zhou2024-682-173}. In \citet{Urquhart2018-473,Urquhart2022-510}, the ATLASGAL clumps at various evolutionary stages encompass all of these physical processes. 
Eventually, a clump forms an embedded cluster with an IMF \citep{Yan2017-607,Zhou2024PASP-1,Zhou2024PASP-2}.  
An ATLASGAL clump associated with an HII region (HII-clump) consists of an embedded cluster surrounded by a gas envelope, with residual gas remaining inside the cluster. High-resolution ALMA (Atacama Large Millimeter/submillimeter Array) observations of HII-clumps presented in \citet{Zhou2024-682-173} reveal distinct gas shells enclosing the protoclusters.
HII-clumps represent the final stage of embedded cluster formation within clumps. They are critical for bridging the stellar kinematics and abundances community with the star formation community, as protoclusters or embedded clusters in HII-clumps mark the earliest stage of gas-free star clusters.  

Following gas expulsion driven by feedback from the embedded cluster, the cluster undergoes expansion \citep{Kuhn2019-870, Wright2024-533, Della2024-683, Zhou2024rm}. Understanding the detailed feedback processes within clumps is essential for tracing the evolutionary pathway from clumps to embedded clusters and ultimately to open clusters, and thus is vital for setting the initial conditions of N-body simulations. The strength of feedback depends on the mass of the embedded cluster, leading to variations in the gas expulsion timescale. More massive clumps give rise to more massive clusters, which, in turn, generate stronger feedback and higher gas expulsion velocities \citep{Dib2013-436,Zhou2024PASP-2}. A correlation is therefore expected between feedback strength, clump (or embedded cluster) mass, and gas expulsion velocity. Low-mass clusters are expected to undergo slower gas expulsion compared to high-mass clusters \citep{Pang2021-912}. 
Moreover, as discussed in Appendix.\ref{nbody}, beyond feedback strength, uncertainties in the star formation efficiency, the geometric shape of the gas shell, and the external gas potential can also be incorporated into the gas expulsion timescale.
Therefore, integrating both hydrodynamics and N-body simulations is crucial for advancing our understanding of early cluster feedback and gas expulsion processes in HII-clumps, as well as for quantifying the impact of the aforementioned factors on the gas expulsion timescale \citep{Sills2018-477,Suin2022-667,Rieder2022-509,Fujii2022-514,Lewis2023-944,Cournoyer2023-521,Rodriguez2023-521,Jo2024-974,Polak2024-690}.

In \citet{expulsion}, we attempted to constrain the gas expulsion mode of initially embedded clusters by inversely analyzing the expansion speed of very young open clusters (<5 Myr) using {\it Gaia} DR3 data. By comparing with N-body simulations, we found that all gas expulsion modes (fast, moderate, slow) remain possible. 
As discussed in Sec.\ref{bhf}, slowly expanding clusters are more likely to retain black holes and black hole multiple systems, making them promising candidates for black hole searches with {\it Gaia} observations.


\subsection{Cluster formation mechanism}

In \citet{Zhou2024-691-293}, we collected samples of Galactic clumps, embedded clusters, and open clusters to compare their physical properties. We showed that the mass distribution of open clusters spans a significantly broader range than that of embedded clusters, by about one order of magnitude. Given the present-day mass distribution of clumps in the Milky Way, the simple evolutionary sequence of a single clump evolving into an embedded cluster and subsequently into an open cluster cannot account for the observed old and massive open clusters. This conclusion is also supported by N-body simulations of individual embedded clusters. 
Furthermore, we compared the separations of open clusters with the typical sizes of molecular clouds in the Milky Way, and found that most molecular clouds are likely to form only one open cluster. Since a molecular cloud generally produces many embedded clusters, this result supports the coalescence scenario. The typical separation between embedded clusters in massive star-forming regions can be $\approx$1~pc \citep{Zhou2024-688L}. Therefore, after expansion, these clusters are expected to undergo coalescence.
There is now extensive literature suggesting that star clusters form through the merging of subclusters, supported by both simulations \citep{Vazquez2009-707,Fujii2012-753,Vazquez2017-467,Howard2018-2,Sills2018-477,Fujii2022-514,Dobbs2022-517,Guszejnov2022-515,Cournoyer2023-521,Polak2024-690} and observations \citep{Sabbi2012-754,Dalessandro2021-909,Pang2022-931,Della2023-674,Fahrion2024-681}. 
Building on the work of \citet{Zhou2024-691-293}, \citet{Zhou2024sub-merger} explored a scenario in which open clusters form via post-gas-expulsion coalescence of embedded clusters within the same parental molecular cloud, and found that the merging of multiple low-mass embedded clusters can reproduce the observed parameter space of open clusters in the Milky Way. 

In the cluster coalescence and individual cluster simulations of \citet{Zhou2024-691-293,Zhou2024sub-merger,Zhou2024sub-mM}, the dynamical states of coalesced and single star clusters differ substantially throughout their evolution, leading to distinct outcomes, as also discussed above. 
Since embedded cluster coalescence is a key pathway for open cluster formation, investigating black hole formation across diverse coalescence scenarios is particularly promising.
In coalescence scenarios, cluster dynamics and black hole formation are influenced by multiple factors, including the mass ratio, spatial separation, relative velocity, collision angle, and initial expansion speed of the subclusters. Possible coalescence scenarios may be inferred from massive star-forming regions in the Milky Way.

Since we adhere to the $m_{\rm max}-M_{\rm ecl}$ relation in all simulations, cluster complexes and single clusters with similar total masses contain significantly different numbers of massive stars, which may in turn has a substantial impact on final black hole production. However,
the $m_{\rm max}-M_{\rm ecl}$ relation has mainly been tested for individual embedded clusters \citep{Weidner2013-434,Zhou2024PASP-2}, and whether it also applies to cluster complexes and their subclusters requires further observational verification.



\section{Conclusion}

The evolution of star clusters is significantly influenced by the gas expulsion mode, with faster expulsion leading to greater mass loss. However, a degeneracy exists whereby clusters of different initial masses (e.g., 1000~M${\odot}$ with slow expulsion and 3000~M${\odot}$ with moderate expulsion) can follow similar evolutionary pathways. In coalescence scenarios, despite subclusters undergoing fast expulsion, the resulting merged cluster can resemble a single cluster with moderate expulsion. This highlights a broader degeneracy between initial cluster mass, gas expulsion timescale, and cluster formation channel (monolithic vs. coalescence).

This degeneracy also plays a key role in black hole retention and dynamical evolution. In individual cluster simulations, slower gas expulsion allows progressively lower-mass clusters to retain central black holes, while in coalescence simulations even a $2207~M_{\odot}$ coalesced cluster can retain one despite fast gas expulsion. 
Variations in gas expulsion mode and formation channel ultimately influence the stellar velocity distribution (within the tidal radius), and thus the expansion speed, which governs both cluster mass loss and black hole retention. 
As the gas expulsion mode transitions from fast to moderate to slow, the fraction of high-velocity stars in the cluster decreases. In contrast, for the cluster complex, despite each subcluster undergoing fast gas expulsion, the velocity distribution of the coalesced cluster is significantly lower overall. The number and configuration of retained black holes (binary or triple systems) depend sensitively on the cluster’s dynamical state, which is itself strongly shaped by the initial gas expulsion mode. Since slowly expanding clusters are more likely to retain black holes and multiple systems, making them prime candidates for black hole searches with {\it Gaia}.

Giant molecular clouds (GMCs) host dense clumps as the primary sites of star formation, evolving through stages that culminate in embedded clusters surrounded by gas envelopes, as seen in ATLASGAL HII-clumps. The subsequent gas expulsion, driven by feedback whose strength correlates with cluster mass, critically determines the cluster's expansion. This process, influenced by factors such as star formation efficiency and gas geometry, sets the initial conditions for N-body simulations. Integrating hydrodynamic and N-body simulations is essential for constraining early feedback and gas expulsion in HII-clumps.

In coalescence scenarios, cluster dynamics and black hole formation are influenced by multiple factors, including the mass ratio, spatial separation, relative velocity, collision angle, and initial expansion speed of the subclusters. Since embedded cluster coalescence is a key pathway for open cluster formation, investigating black hole formation across diverse coalescence scenarios is particularly promising. 
Adherence to the $m_{\rm max}-M_{\rm ecl}$ relation in both individual and coalescence simulations implies that cluster complexes and single clusters of similar total mass contain different numbers of massive stars, which may significantly impact final black hole production, although the applicability of this relation to cluster complexes and their subclusters requires further observational verification.

\begin{acknowledgements}
Thanks to the referee for the detailed and constructive comments, which have significantly contributed to improving this work.
\end{acknowledgements}

\bibliographystyle{aa} 
\bibliography{ref}


\appendix

\section{N-body simulations}\label{nbody}

\subsection{Individual cluster simulation}\label{individual}

The parameters for the simulation are summarized from previous works (i.e.,
\citealt{Kroupa2001-321,
Baumgardt2007-380,
Banerjee2012-746,
Banerjee2013-764,
Banerjee2014-787,
Banerjee2015-447,
Oh2015-805,
Oh2016-590,
Banerjee2017-597,
Brinkmann2017-600,
Oh2018-481,
Wang2019-484,
Pavlik2019-626,
Dinnbier2022-660,
Zhou2024-691-204,
Zhou2024-691-293}).
The previous simulations have already demonstrated the effectiveness and rationality of the parameter settings (see below). The influence of different parameter settings on simulation results and the discussion of the multidimensional parameter space can also be found in the works cited above. 

The initial density profile of all clusters is the Plummer profile \citep{Aarseth1974-37, HeggieHut2003, Kroupa2008-760}, an appropriate choice since the molecular clouds’ filaments in which stars form have been found to have Plummer-like cross sections \citep{Malinen2012-544,Andre2022-667},
and open star clusters can also be described by the Plummer model \citep{Roeser2011-531,Roeser2019-627}.
Moreover, such a specific initial profile does not significantly affect the overall expansion rate of a cluster, as discussed in \citet{Banerjee2017-597}, 
which is primarily governed by the total stellar mass loss and the dynamical interactions occurring within the inner part of the cluster. 
The half-mass radius, $r_{h}$, of the cluster is given by the $r_{\rm h}-M_{\rm ecl}$ relation \citep{Marks2012-543}:
\begin{equation}
 \frac{r_{\rm h}}{{\rm pc}}=0.10_{-0.04}^{+0.07}\times\left(\frac{M_{\rm ecl}}{M_{\odot}}\right)^{0.13\pm0.04}\;.
 \label{rm}
\end{equation}
All clusters are fully mass segregated ($S$=1), with no fractalization, and in a state of virial equilibrium ($Q$=0.5). $S$ and $Q$ are the degree of mass segregation and the virial ratio of the cluster, respectively. More details can be found in \citet{Kupper2011-417} and the user manual for the \texttt{McLuster} code.
The initial segregated state is detected for young clusters and star-forming clumps/clouds \citep{Littlefair2003-345,Chen2007-134,Portegies2010-48,Kirk2011-727,Getman2014-787,Lane2016-833,
Alfaro2018-478,Plunkett2018-615, Pavlik2019-626, Nony2021-645, Zhang2022-936,
Xu2024-270}, but the degree of mass segregation is not clear. 
In simulations of the very young massive clusters R136 and NGC 3603 with gas expulsion by \citet{Banerjee2013-764}, mass segregation does not influence
the results. In \citet{Zhou2024-691-204}, we compared $S$=1 (fully mass segregated) and $S$=0.5 (partly mass segregated) and found similar results. We also discussed settings with and without fractalization in \citet{Zhou2024-691-204}; the results of the two are also consistent.
The initial mass functions of the clusters are chosen to be canonical \citep{Kroupa2001-322}, with the most massive star following the $m_{\rm max}-M_{\rm ecl}$ relation  of \citet{Weidner2013-434}:
\begin{equation}
\label{eq:mmaxmecl2}
y = a_0 + a_1 x + a_2  x^2 + a_3  x^3 ,
\end{equation}
where $y$ = $\log_{10}(m_\mathrm{max}/M_\odot)$, $x$ = $\log_{10}(M_\mathrm{ecl}/M_\odot)$, $a_0$ = -0.66, $a_1$ = 1.08, $a_2$ = -0.150, and $a_3$ = 0.0084. We assumed the clusters to be at solar metallicity \citep[i.e., $Z=0.02$;][]{von2016ApJ...816...13V}. 
The clusters travel along circular orbits within the Galaxy, positioned at a galactocentric distance of 8.5 kpc, and are moving at a speed of 220 km s$^{-1}$.

The initial binary setup follows the method described in \citet{Wang2019-484}.
All stars are initially in binaries, that is to say, $f_{\rm b}$=1, where $f_{\rm b}$ is the primordial binary fraction.
\citet{Kroupa1995-277-1491,Kroupa1995-277-1507} propose that stars with masses below a few $M_\odot$ are initially formed with a universal binary distribution function and that star clusters start with a $100\%$ binary fraction. Inverse dynamical population synthesis was employed to derive the initial distributions of binary periods and mass ratios. \cite{Belloni2017-471} introduces an updated model of pre-main-sequence eigenevolution, originally developed by \cite{Kroupa1995-277-1507}, to account for the observed correlations between the mass ratio, period, and eccentricity in short-period systems. For low-mass binaries, we adopted the formalism developed in \citet{Kroupa1995-277-1491,Kroupa1995-277-1507} and \citet{Belloni2017-471} to characterize the period, mass ratio, and  eccentricity distributions.
For high-mass binaries (OB stars with masses $>5$~M$_\odot$), we utilized the \citet{Sana2012-337} distribution, which is derived from O stars in OCs. The distribution functions of the period, mass ratio, and eccentricity are presented in \cite{Oh2015-805} and \cite{Belloni2017-471}. 

Accurately modeling gas removal from embedded clusters is challenging due to the complexity of radiation hydrodynamical processes, which involve uncertain and intricate physical mechanisms. 
To simplify the approach, we simulated the key dynamical effects of gas expulsion by applying a diluting, spherically symmetric external gravitational potential to a model cluster, following the method presented in \citet{Kroupa2001-321} and \citet{Banerjee2013-764}. 
This analytical approach is partially validated by \citet{Geyer2001-323}, who conducted comparison simulations using the smoothed particle hydrodynamics method to treat the gas. The hydrodynamics+N-body simulations in \citet{Farias2024-527} also confirm that the exponential decay model presented in equation.\ref{eq:mdecay} generally provides a good description of gas removal driven by radiation and wind feedback.
Specifically, we used the potential
of the spherically symmetric, time-varying mass distribution:
\begin{eqnarray}
M_g(t)=& M_g(0) & t \leq \tau_d,\nonumber\\
M_g(t)=& M_g(0)\exp{\left(-\frac{(t-\tau_d)}{\tau_g}\right)} & t > \tau_d.
\label{eq:mdecay}
\end{eqnarray}
Here, $M_g(t)$ is the total mass of the gas; it is spatially distributed with the Plummer density distribution \citep{Kroupa2008-760} and starts depleting after a delay of $\tau_d$, and is totally depleted 
within a timescale of $\tau_g$. The Plummer radius of the gas distribution is kept time-invariant \citep{Kroupa2001-321}.
This assumption is an approximate model of the effective gas reduction within the cluster in the situation that gas is blown out while new gas is also accreting into the cluster along filaments such that the gas mass ends up being reduced with time but the radius of the gas distribution remains roughly constant. As discussed in \citet{Urquhart2022-510},
the mass and radius distributions of the ATLASGAL clumps at different evolutionary stages are quite comparable.
We used an average velocity of the outflowing gas of $v_g\approx10$ km s$^{-1}$, which is the typical sound speed in an HII region. This gives
$\tau_g=r_h(0)/v_g$,
where $r_h(0)$ is the initial half-mass radius of the cluster. As for the delay time, we take the representative value of $\tau_d\approx0.6$ Myr
\citep{Kroupa2001-321}, this being about the lifetime of the ultracompact HII region. 
As shown in \citet{Banerjee2013-764}, varying the delay time, $\tau_d$, primarily results in a temporal shift in the cluster’s rapid expansion phase, without significantly impacting its subsequent evolution for times greater than $\tau_d$. 
Protoclusters typically form in hub-filament systems \citep{Motte2018-56,Vazquez2019-490, Kumar2020-642,Zhou2022-514}, which are located in hub regions. Compared to the surrounding filamentary gas structures, the hub region, as the center of gravitational collapse, is usually more regular, as shown in \citet{Zhou2022-514,Zhou2024-686-146}. 
Thus, modeling a spherically symmetric mass distribution is appropriate.

In this work, we assumed a SFE $\approx$ 0.33 as a benchmark (i.e., $M_{g}(0)$ = 2$M_{\rm ecl}(0))$. This value has been widely used in the simulations cited above and has been proven effective in reproducing the observational properties of stellar clusters. Such a SFE is also consistent with the value obtained from hydrodynamical calculations that include self-regulation \citep{Machida2012-421,Bate2014-437} and as well with observations of embedded systems in the solar neighborhood \citep{Lada2003-41,Megeath2016-151}.
In \citet{Zhou2024-688L}, by comparing the mass functions of the ATLASGAL clumps and the identified embedded clusters, we found that a SFE of $\approx$ 0.33 is typical for the ATLASGAL clumps. 
In \citet{Zhou2024PASP-1}, assuming SFE = 0.33, it was shown that the total bolometric luminosity of synthetic embedded clusters can fit the observed bolometric luminosity of ATLASGAL clumps with HII regions. In \citet{Zhou2024PASP-2}, we directly calculated the SFE of ATLASGAL clumps with HII regions and found a median value of $\approx$0.3.


Embedded clusters form in clumps. More massive clumps can produce more massive clusters, leading to stronger feedback and a higher gas expulsion velocity \citep{Dib2013-436}. 
There should be correlations between the feedback strength, the clump (or cluster) mass, and the gas expulsion velocity ($v_g$). And low-mass clusters should have a slower gas expulsion process compared with high-mass clusters. As shown in \citet{Pang2021-912}, low-mass clusters tend to agree with the simulations of slow gas expulsion. Except for the feedback strength, the SFE determines the total amount of the remaining gas, which also influences the timescale of gas expulsion. The gas expulsion process is driven by feedback, and the effectiveness of the feedback will depend on the geometric shape of the gas shell surrounding the embedded cluster  \citep{Wunsch2010-407,Rahner2017-470}. Therefore, a complex or nonspherical gas distribution would also change the timescale of gas expulsion. Moreover,
the total amount of the residual gas not only affects the gas expulsion timescale, but also significantly influences the strength of the gas potential. The strength of the external gas potential may have a considerable impact on the evolution of embedded star clusters. The total amount of residual gas is determined by the SFE of the clump. As verified in \citet{Zhou2024-691-293}, a lower SFE is equivalent to a shorter gas expulsion timescale. 
In short, the uncertainty of the above parameters can ultimately be incorporated into the timescale of gas expulsion. Thus, apart from the fast gas expulsion with the gas decay time, $\tau_g$, described above, we also simulated slow and moderate gas expulsions. 
For the slow gas expulsion, the gas decay time was set to 10$\tau_g$ \citep{Zhou2024-691-293}. 
The moderate gas expulsion is between the fast and slow gas expulsions. Considering the large parameter space between $\tau_g$ and 10$\tau_g$, we simulated two kinds of moderate gas expulsions: 2$\tau_g$ ("moderate2" or "mod2") and 5$\tau_g$ ("moderate1" or "mod1"). 

\subsection{Procedure}

The \texttt{McLuster} program \citep{Kupper2011-417} was used to set the initial configurations. 
The dynamical evolution of the model clusters was computed using the state-of-the-art \texttt{PeTar} code \citep{Wang2020-497}. 
\texttt{PeTar} employs well-tested analytical single and binary stellar evolution recipes (SSE/BSE)
\citep{Hurley2000-315,Hurley2002-329,Giacobbo2018-474,Tanikawa2020-495,Banerjee2020-639}. 

\subsection{Cluster coalescence simulation}\label{coalescence}

For the massive star-forming region  NGC 6334,
we have identified its subclusters using the dendrogram algorithm according to the surface density distributions of stars in \citet{Zhou2024-688L}, and calculated their physical parameters. Considering that
the mass-radius relation of embedded clusters can be well fitted by the $\approx$1 Myr expanding line in \citet{Zhou2024-691-204}, 
we simulated the evolution of the subclusters in this MSFR for the first 1 Myr using the recipe described in Sec.\ref{individual}. The initial masses of the subclusters strictly follow the observed values, and each subcluster is simulated individually.
Then, all subclusters were collected together to create an initial configuration similar to the observation.

We first assume all embedded clusters are initially at rest and located on the same plane ("fast"). A more realistic scenario is that clusters have relative velocities and line-of-sight spatial separations ("fast-vd"). To take these two factors into account, we approximate the molecular cloud containing the clusters as an ellipsoid. More details can refer to \citet{Zhou2025-537-845}.
Then we randomly distribute clusters along the Z-axis (line-of-sight). 
As discussed in \citet{Zhou2025-537-845}, we take the velocity dispersion of the original molecular cloud as 2 km s$^{-1}$ and assume that the subclusters inherit the velocity dispersion of the cloud. The system's center has a velocity of 0 km s$^{-1}$, and the velocity of the outermost subcluster is 2 km s$^{-1}$. The velocities of other subclusters are distributed according to the Larson relation, i.e. $v \propto L^{0.5}$, where $L$ is the distance to the system's center.

\end{document}